\documentclass[12pt]{iopart}

\usepackage{amssymb}
\usepackage{amsfonts}
\usepackage{bm}
\usepackage{epsfig}
\usepackage{epsf}                          
\usepackage[]{graphicx}  

\usepackage{color}

\renewcommand {\d} {{\rm d}}
\renewcommand {\i} {{\rm i}}

\newcommand {\ee}  {{\rm e}}

\newcommand {\E}  {{\varepsilon}}
\newcommand {\om}  {{\omega}}
\newcommand {\Om}  {{\Omega}}

\newcommand {\kp}  {{\kappa}}
\newcommand {\kpd}  {{\kappa_{\rm d}}}
\newcommand {\kpa}  {{\kappa_{\rm a}}}

\newcommand {\Ld} {{L_{\rm d}}}
\newcommand {\La} {{L_{\rm a}}}

\newcommand {\ymax} {{y_{\max}}}

\newcommand {\amax} {{a_{\max}}}
\newcommand {\amin} {{a_{\min}}}

\newcommand {\pmax} {{p_{\max}}}
\newcommand {\pmin} {{p_{\min}}}

\newcommand {\calA}  {{\cal A}}
\newcommand {\calC}  {{\cal C}}
\newcommand {\calD}  {{\cal D}}
\newcommand {\calF}  {{\cal F}}
\newcommand {\calI}  {{\cal I}}
\newcommand {\calJ}  {{\cal J}}
\newcommand {\calS}  {{\cal S}}

\newcommand {\dUmax} {U^{\prime}_{\max}}

\newcommand {\bfs}  {{\bf s}}
\newcommand {\bfu}  {{\bf u}}

\begin{document}
\jl{2}


\title[]{The influence of the structure imperfectness of a crystalline
undulator on the emission spectrum}

\author{
A Kostyuk$^{1}$, 
A V Korol$^{1,2}$, 
A V Solov'yov$^{1}$ 
and 
Walter Greiner$^{1}$
}

\address{
$^1$ Frankfurt Institute for Advanced Studies, Johann Wolfgang 
G\"othe-Universit\"at,
Ruth-Moufang-Str. 1, 60438 
Frankfurt am Main, Germany}
\address{
$^2$ Department of Physics, St. Petersburg State Maritime 
Technical University, Leninskii prospect 101,
St. Petersburg 198262, Russia}
\ead{kostyuk@th.physik.uni-frankfurt.de, a.korol@fias.uni-frankfurt.de, 
solovyov@fias.uni-frankfurt.de}


\begin{abstract}
We study the influence of an imperfect structure of a crystalline undulator
on the spectrum of the undulator radiation.
The main attention is paid to the undulators in which the periodic bending
in the bulk appears as a result of a regular (periodic) surface deformations.
We demonstrate that this method of preparation of a crystalline
undulator inevitably leads to a  variation of the bending 
amplitude over the crystal thickness and to the presence of the
subharmonics with smaller bending period.
Both of these features noticeably influence the monochromatic pattern of the
undulator radiation.
\end{abstract}

\section{Introduction \label{Introduction}}

In this paper we present the qualitative and quantitative analysis 
of the influence of imperfect structure of a crystalline undulator
(in particular, the influence of the variation of the bending amplitude 
over the crystal thickness due to the stress applied to its surface) 
on the spectral distribution of the radiation.
The parameters of crystalline undulators (including the types and lengths 
of crystals, the periods of bending, the positron energies, and 
the range of photon energies) as well as the methods of preparation of 
periodically bent structures discussed below in the paper
correspond to those which are available for the experiments 
to be carried out within the PECU project\cite{PECU}.  

A periodically bent crystal together with a bunch of ultra-relativistic 
charged particles which undergo planar channeling constitute a 
crystalline undulator.
In such a system there appears, in addition to the well-known channeling 
radiation, the  undulator type radiation which is due to the periodic 
motion of channeling particles which follow 
the bending of the crystallographic planes\cite{KSG1998,KSG1999}. 
The intensity and characteristic frequencies of this 
radiation can be varied by changing the beam energy 
and the parameters of bending.
In the cited papers as well as in the subsequent publications
(see the review Ref.~\cite{KSG2004_review} and the references therein)
a feasibility was proven to create a short-wave crystalline undulator 
that will emit intensive monochromatic radiation when a pulse of 
ultra-relativistic  positrons is passed through its channels.
More recently, it was demonstrated \cite{TKSG2007} that 
the undulator based on the electron channeling is also feasible.
A number of corresponding novel numerical results were presented to
illustrate the developed theory, including, in particular, the
calculation of the spectral and angular characteristics of the new
type of radiation.

\begin{figure}[ht]
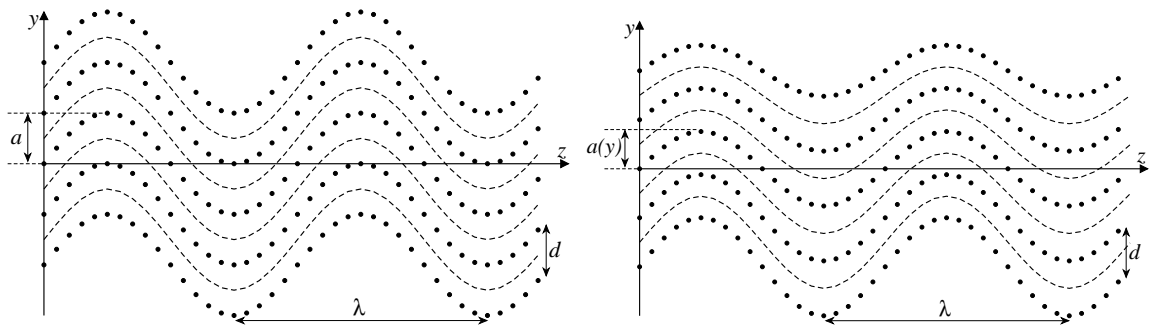

\parbox{15.5cm}
{
\includegraphics[clip,width=7.5cm,angle=0]{figure1a.eps}
\hspace*{0.2cm}%
\includegraphics[clip,width=7.5cm,angle=0]{figure1b.eps}
}
\caption
{Schematic representation of a crystalline undulator 
with a constant bending amplitude $a$ (left panel) and
with a varied amplitude $a(y)$ (right panel).
Circles denote the atoms belonging to neighbouring
crystallographic planes (separated by the distance $d$) which are
periodically bent with a period $\lambda$. 
The dashed curves denote the centerlines of the planar channels.
} 
\label{figure.1}
\end{figure}

Although the operational principle of a crystalline undulator does not depend
on the type of a projectile below in this paper we will consider the
case of a positron channeling. 
Under certain conditions \cite{KSG1998,KSG1999}, an
ultra-relativistic positron, which enters the crystal at
the angle smaller than the Lindhard critical angle \cite{Lindhard}, 
will penetrate through the crystal following the bendings of its planes.
Provided the bending amplitude $a$ greatly exceeds the interplanar distance
$d$ (see figure \ref{figure.1} left) one can disregard 
the oscillations due to the action of the interplanar force, --
the channeling oscillations\cite{Lindhard}.
In this case the trajectory of the particle can be associated with the 
periodic profile of channel centerline.
The undulator radiation appears as a result of this periodic motion of the
particle.
Thus, the operational principle of a
crystalline undulator is the same as for a conventional
one\cite{Alferov1989,RullhusenArtruDhez}, in which 
the monochromaticity of the radiation is the result of a constructive 
interference of the photons emitted from similar parts of trajectory. 
 
Usually, when discussing the properties of a crystalline undulator and 
of its radiation,  one considers the case
of {\it a perfect} crystalline undulator.
By this term we will understand the crystal whose planes are bent 
periodically following a perfect harmonic shape, 
$y(z) = a\sin (2\pi z/\lambda)$, see  figure \ref{figure.1} (left).
For clarity, let us stress that we consider the case when 
the quantities $d$, $a$ and $\lambda$  satisfy the double 
inequality $d \ll a \ll \lambda$.
Typically, $d\sim 1$ \AA, $a=10\dots10^2 d$ and 
$\lambda \sim 10^{-5}\dots10^{-4} a$.
The spectral-angular distribution from the perfect undulator
is characterized by a specific pattern which implies that for 
each value of the emission angle the spectrum consists of a set of narrow, 
well-separated and powerful peaks corresponding to different harmonics of 
radiation.
In principle, the perfect crystalline undulators can be 
produced by using the technologies of growing Si$_{1-x}$Ge$_x$ 
structures \cite{Breese97}.
In this case, by varying the Ge content $x$ one can 
obtain periodically bent crystalline structure
\cite{MikkelsenUggerhoj2000,Darmstadt01}.
The technological restrictions imposed by this method on the
crystalline undulator length is $L\leq 140\dots150\ \mu$m. 
 
The periodic bending can also be achieved by making regularly spaced 
grooves on the crystal surface 
either by using a diamond blade \cite{BellucciEtal2003,GuidiEtAl_2005}
or by means of laser-ablation.
The latter method was used \cite{Uggerhoj2006_Connell2006}
to prepare the Si-based crystalline undulators for the PECU experiments.
In either case, the regular surface deformation results in the 
periodic pattern of the crystallographic planes bending in the bulk.
The question which appears in connection with these methods of preparation 
concerns the quality of the periodically bending.
Indeed, for a crystal of a finite thickness it is natural to expect that
the surface deformations, regularly spaced with the period $\lambda$, 
result in the volume deformations of the same period but of a varied
amplitude of bending, $a=a(y)$ (see figure \ref{figure.1} (right)). 
The latter has maximum value in the surface layer but decreases
with the penetration distance.
Therefore, it is important to carry out a quantitative analysis 
(a) of the structure of this {\it imperfect} periodic bending in the bulk,
 and 
(b) of its influence on the spectrum of undulator radiation. 
Both of these problems constitute the subject of the present paper.

The paper is organized as follows. 
In section \ref{dE} we briefly outline the basic formulae and 
definitions which refer to a perfect crystalline undulator.
Section \ref{Averaging} describes, in general terms,  
the modifications to be introduced to the formalism due to the 
imperfectness of the crystalline undulator.
In section  \ref{Formalism} we present the formalism and carry out 
numerical analysis of the periodic deformations in the bulk caused by a 
regular stress applied to the crystal surface.
Finally, in section \ref{AveragedSpectra} we carry out quantitative
analysis of the differences in the
emission spectra formed in a perfect undulator and in the undulator
created by means of periodic surface deformations.

\section{Spectral-angular distribution of the radiation 
from a perfect crystalline undulator \label{dE}}
The spectral distribution of the  energy $E$ of radiation emitted by an 
ultra-relativistic ($v\approx c$) positron in a perfect crystalline undulator 
can be written in the following form \cite{Dechan01,SPIE1}:
\begin{eqnarray}
{\d^3 E \over \hbar \d\omega\,\d\Omega }
=
S(\om,\theta,\varphi)\,
 \calD_N(\eta,\kpd,\kpa)\, ,
\label{dE.1}
\end{eqnarray}
where $\theta\ll 1$ and $\varphi$ are the emission angles 
with respect to the undulator axis chosen along $z$,
$\d\Omega = \theta \d \theta \d \varphi$ is the emission solid angle.
The factor $\calD_N(\eta,\kpd,\kpa)$ is explained further in this section.
The function $S(\om,\theta,\varphi)$ is given by
\begin{eqnarray}
\fl
S(\om,\theta,\varphi)
=
{\alpha \over 4\pi^2}\,
{\omega^2 \over \gamma^2\omega_0^2}
\left\{
p^2\left|I_1\right|^2
+
\gamma^2\theta^2 \left|I_0\right|^2
-
2p\gamma\, \theta \cos \varphi\, {\rm Re}\left(I_0^{*} I_1\right) 
\right\}\, ,
\label{dE.2}\\
\fl
 I_m
=
\int_0^{2\pi}
\d \psi\,
\cos^m\psi
\exp\Biggl(
\i \left[\eta \psi 
+ {p^2\omega \over 8\gamma^2\omega_0} \sin(2\psi) 
- {p \omega \over \gamma\omega_0} \theta\cos\varphi\sin \psi
\right]
\Biggr),
\quad
m=0,1\,.
\label{dE.3}
\end{eqnarray}
Here $\alpha \approx 1/137$ is the fine structure constant, 
$\gamma=\E/mc^2$ is the relativistic Lorentz factor with $m$ standing for the
positron mass and $\E$ for its energy,
$\om_0= 2\pi c/\lambda$ and
$p$ is the undulator parameter,
\begin{equation}
p = 2\pi \gamma {a\over \lambda}.
\label{No.3} 
\end{equation}
The  parameter $\eta$ is defined as follows
 \begin{eqnarray}
\eta = {\omega \over 2 \gamma^2\omega_0}\,
\left(1 + \gamma^2\theta^2 + {p^2 \over 2}
\right).
\label{dE.4}
\end{eqnarray}
A peculiar feature of the  undulator radiation is that for each value 
of the emission angle $\theta$ the spectral distribution consists of 
a set of narrow and equally spaced peaks
(harmonics) the frequencies $\om_j$ one defines letting 
$\eta = j = 1,2,3\dots$:
 \begin{eqnarray}
\om_j = {4 \gamma^2\omega_0 \over p^2 + 2+ 2\gamma^2\theta^2}\, j.
\label{dE.4a}
\end{eqnarray}
The widths $\Delta\om_j$  of the peaks satisfy the inequality
$\Delta\om_j \ll \om_j$, so that all peaks are well separated.

In an ideal undulator
(i.e., in which positrons and photons propagate in 
vacuum)\footnote{The term 'ideal undulator' must not be 
mixed up with the term 'perfect undulator', which is used throughout 
the paper and stands for the crystalline
undulator with a perfect harmonic pattern of periodically bent channels.} 
the peak intensity is proportional to the 
squared number of periods.
Formally, it follows from the fact that $\d^3 E$ is proportional to 
$D_N(\eta)\equiv\Bigl(\sin (N\pi\eta)/ \sin(\pi\eta)  \Bigr)^2$ 
which behaves as $N^2$ for integer $\eta$ (e.g., \cite{Alferov1989}).
This factor reflects the constructive interference of radiation emitted
from each of the undulator periods.
Consequently, in an ideal undulator one can increase unrestrictedly 
the radiated intensity by increasing of the undulator length $L=N\lambda$.

The situation is different for a crystalline undulator, where 
the number of channeling particles and the number of 
photons which can emerge from the crystal decrease with the growth of $L$.
In Refs. \cite{Dechan01,SPIE1} the quantitative study of the influence of 
the dechanneling and the photon attenuation on the spectral-angular
distribution was presented.
The main result of this analysis is that the peak intensity is no longer 
proportional to $N^2$. 
It was shown, that in a crystalline undulator the factor $D_N(\eta)$ must 
be substituted with $\calD_N(\eta,\kpd,\kpa)$, which depends not only on $N$ 
and $\eta$ but also on the ratios $\kpd ={L/\Ld}$ and $\kpa ={L/ \La}$.
Here $\Ld$ stands for the dechanneling length which is the
mean penetration distance covered by a channeling particle.
The quantity $\La$, called the attenuation length, defines 
the scale within which the intensity of a photon flux propagating 
through a crystal decreases by a factor of $e$  due to the 
processes of absorption and scattering.
A convenient formula for $\calD_N(\eta,\kpd,\kpa)$, which enters the right-hand side
of (\ref{dE.1}), is as follows \cite{SPIE1}:
\begin{eqnarray}
\fl
\calD_N(\eta,\kpd,\kpa)
&=
{4N^2 \over \kp_{\rm a}^2 + 16N^2\sin^2\pi(\eta-j_{\eta})}
\Biggl[
{\kpa\,\ee^{-\kpd} \over \kpa -\kpd}
-
{2\kpd-\kpa\over \kpa -\kpd}
{(\kp_{\rm a}^2+4\phi^2)\, \ee^{-\kpa}
\over(2\kpd-\kpa)^2 +4\phi^2}
\nonumber\\
\fl
&\quad
-
2
\left(
\cos\phi
+
2\kpd\,
{
2\phi\,\sin\phi-(2\kpd-\kpa)\cos\phi
\over
(2\kpd-\kp_{\rm a})^2 +4\phi^2}
\right)
\ee^{-(2\kpd+\kpa)/2}
\Biggr]\, ,
\label{dE.5}
\end{eqnarray}
where $\phi=2\pi(\eta-j_{\eta}) N$ (with $j_{\eta}$ being the closest 
positive integer  to $\eta$).

Despite a cumbersome form of the right-hand side of 
(\ref{dE.5}) its main features can be easily understood. 
The most important is that, as in the case of an ideal undulator
(to which $\calD_N(\eta,\kpd,\kpa)$ reduces in the limit $\Ld = \La = \infty$)
the main maxima of $\calD_N(\eta,\kpd,\kpa)$  correspond to $\eta$ integers, 
and, therefore, the harmonic frequencies are still defined by
(\ref{dE.4}) with $\eta=j$.
For finite $\Ld$ and $\La$, the maximum value of   
$\calD_N(\eta,\kpd,\kpa)$ is smaller than $N^2$ whereas the width of the 
peak is larger than that in the corresponding ideal undulator.

The attenuation length depends on the photon energy,  $\La = \La(\om)$,
and can be calculated as the inverse mass attenuation coefficient 
which are  tabulated for all elements and for a wide range of photon 
frequencies \cite{Hubbel,ParticleDataGroup2006}.

The dechanneling effect stands for a gradual increase in the 
transverse energy of a channeled particle due to inelastic collisions 
with the crystal constituents \cite{Lindhard}.
At some point the particle gains a transverse energy 
higher than the planar potential barrier and leaves the channel.
In a straight crystal dechanneling length $\Ld$ of a positron depends 
on the crystal and on the energy of the projectile.
In a bent crystal the potential barrier changes due to the centrifugal force, 
and, as a result,  $\Ld$ becomes dependent on the curvature of the channel.
A stable channeling of a projectile in a periodically bent channel occurs 
if the maximal centrifugal force in the channel $F_{\rm cf}$ is less 
than the  maximal interplanar force $\dUmax$, 
i.e. $C\equiv F_{\rm cf}/\dUmax < 1$.
For an ultra-relativistic particle $F_{\rm cf}\approx \E/R_{\min}$, 
where $R_{\min}$ is the minimum curvature radius 
of the bent channel.
In a perfect crystalline undulator $R_{\min}=\lambda^{2}/4\pi^{2}a$, 
therefore, the condition for a stable channeling reads \cite{KSG1998}:
\begin{eqnarray}
C ={4\pi^2\E a\over  \dUmax \lambda^2} < 1.
\label{dE.6}
\end{eqnarray}
Using the diffusion model \cite{BiryukovChesnokovKotovBook}
one can be demonstrated that dechanneling length in a periodically bent 
crystal becomes dependent on the parameter $C$.
In the case of ultra-relativistic positrons the expression for  
$\Ld\equiv \Ld(C)$ can be written as follows \cite{KSG1999,KSG2001_Dech}:
\begin{equation}
\Ld(C)=(1-C)^{2} \Ld(0), 
\label{dE.7}
\end{equation}
where $\Ld(0)$ is the dechanneling length a straight channel.
This quantity can be estimated as
$\Ld(0)=(256/9\pi^2)(a_{\rm TF}\, d \gamma/r_0\,\Lambda_{\rm c})$ 
\cite{KSG1999,BiryukovChesnokovKotovBook}, where $r_0$ and $a_{\rm TF}$ are 
the classical radius of the electron and the Thomas-Fermi radius of the
crystal atom, and $\Lambda_{\rm c}=\ln(\sqrt{2 \gamma} \, mc^{2}/I)-23/24$
($I$ stands for an average ionization potential of the atom).

Apart from the dechanneling length, the bending parameter $C$ also defines
another important quantity, {\em the acceptance} of the crystal.
The acceptance represents by itself the fraction of the particles
trapped in the channeling mode 
(or, in other words, it is equal to the ratio of the
phase volume corresponding to the channeling regime to the phase volume 
of the beam particles at the entrance of the crystal).
In comparison with a linear crystal, the acceptance of the bent crystal
gains additional decrease due to the presence of the centrifugal force in the 
channels \cite{BiryukovChesnokovKotovBook}.
Generalizing the result of Ref. \cite{KaplinVorobiev1978}, one 
writes the following expression for the acceptance $\calA(C)$ of the 
periodically bent channel and for the case of a parallel beam:
\begin{equation}
\calA(C) = (1-C) \calA(0), 
\label{dE.8}
\end{equation}
where $\calA(0)$ is the acceptance of the linear crystal and $C\leq 1$.
If the centrifugal force exceeds the interplanar force (i.e., $C>1$) then
one uses the identity $\calA(C)\equiv 0$, which means that the channel does 
capture the particles. 

In connection with the radiation from a crystalline undulator, the
acceptance comes into play when one calculates the 
energy emitted per particle of the bunch. 
To obtain this quantity in a perfect undulator, 
one multiplies the right-hand side of (\ref{dE.1}) by the acceptance
$\calA(C)$.

The formulae (\ref{dE.1})-(\ref{dE.8}) allow one to carry out 
numerical analysis of the spectral-angular distribution
formed in a perfect crystalline undulator.
In a recent paper \cite{SPIE2007} the results of such analysis were reported 
for 0.6 and 10 GeV positrons channeling
through periodically bent Si and Si$_{1-x}$Ge$_x$ crystals.

\section{Emission from the undulator with a varied bending amplitude
\label{Averaging}}

In this section we analyze the modifications in the formalism of 
spectral-angular distribution of the radiation from a crystalline undulator 
which appear due to the imperfectness 
of periodic bending of the crystallographic planes.

For the sake of clarity let us discuss the geometrical conventions 
which will be used below.
Figures \ref{figure.1} and \ref{figure.2} help a reader to follow the text.
We assume that the crystal has the form of a rectangular box.
Its length, $L$, width, $l$,  and thickness, 
$h$, are measured along the $z$, $x$ and $y$ directions, respectively. 
We chose the frame in which the values $\pm h/2$ denote 
the $y$-coordinates of the upper and lower surfaces of the crystal.
Hence, it is assumed that $y=0$ labels the central $(xy)$-plane of the crystal
(the midplane).
Initially non-deformed crystallographic planes are perpendicular to the
 $y$ axis.
We assume that periodic deformation of the crystalline structure 
occurs only in the $(yz)$-plane, so that there is  
no deformation in the $x$-direction.
In the deformed crystal the bunch of channeling particles propagates
in the $(yz)$-plane along $z$ direction.
With $\sigma_y$ we denote the bunch size along the $y$ direction.
The width of the bunch (i.e., its size in $x$) is not used in our model.

Suppose that bending amplitude is not constant over the crystal thickness
but changes according to a law $a=a(y)$, as illustrated
by figure \ref{figure.1}.
The particular form of this dependence is discussed in Section 
\ref{Formalism} where we analyze the deformation of the crystal
interior due to the periodic stress applied to its surfaces.
At the moment, the only restrictions implied on the dependence $a(y)$ are:
(a) the change in amplitude on the scale of interplanar spacing $d$ 
is negligible;
(b)  a strong inequality $a(y)\ll \lambda$ is valid for all $y$.

Particles in the bunch are randomly distributed along the $y$-axis.
Therefore, we may assume that at the entrance a particle 
can be captured in any channel located within the 
interval $y=[-\ymax,\ymax]$, where $\ymax$ stands for the smallest from 
$h/2$ and $\sigma_y/2$. 
Being captured in the channel at some $y$ point, the particle undulates
and emits the radiation corresponding to the undulator with the 
amplitude $a(y)$.
For a fixed value of the period $\lambda$, the amplitude  
defines two important quantities which are the undulator parameter $p$ and the 
bending parameter $C$.
The latter, in turn, defines the  dechanneling length and the acceptance
(see eqs. (\ref{dE.7}) and (\ref{dE.8})).
The influence of $p$ and $\Ld$ on the spectral-angular distribution of radiation
is discussed in detail further in this section.
Here we note that the acceptance $\calA(C)$ determines the probability of a 
particle to be captured in the channeling more.
If $C=C(y)$,  this probability depends on the entrance coordinate $y$.
Therefore, to obtain the distribution (per particle) of radiation formed in 
the crystal with varied amplitude of bending 
one should (a) multiply eq. (\ref{dE.1}) by the acceptance (\ref{dE.8}), and
(b) carry out the averaging over the interval $y=[-\ymax,\ymax]$.
This leads to the formula: 
\begin{eqnarray}
\left\langle {\d^3 E \over \hbar\, \d\om\,\d\Om}\right\rangle
=
{1\over 2\ymax}
\int_{-\ymax}^{\ymax}   
\calA(C)\,{\d^3 E(y) \over \hbar\, \d\om\,\d\Om} \,\d y
\label{Averaging.1}
\end{eqnarray}
Here $\d^3 E(y) / \hbar\, \d\om\,\d\Om$ stands for the  
the distribution (\ref{dE.1}) obtained for the amplitude
$a(y)$.
The right-hand side of this equation represents the spectral-angular
distribution per particle averaged over the width of the crystal 
(or of the bunch if $\sigma_y<h$)\footnote{The contribution to the integral
comes only from the regions where $C(y)<1$, 
otherwise $\calA(C)\equiv 0$,
see eq. (\ref{dE.8}). }.
If the amplitude does not change within the interval of integration, 
the formula (\ref{Averaging.1}) reduces to eq. (\ref{dE.1}) multiplied by 
the acceptance corresponding to the fixed $C$-value.

Let us discuss in more detail the influence of the dependence $a(y)$ 
on the characteristics of the crystalline undulator radiation.

It is clear that for the $y$-varying amplitude the terms, 
dependent on $p$ in the function $S(\om,\theta,\varphi)$
(see (\ref{dE.2})),
change with $y$ since $p\propto a(y)$ (see (\ref{No.3})).
Apart from this influence, the proportionality can lead to the variation 
of the harmonics frequencies $\om_j$ and, consequently, to the loss of the
monochromaticity of the radiation. 
Indeed, let $\amin$ and $\amax$ denote the minimum and maximum
amplitudes within the interval $[-\ymax,\ymax]$.
The corresponding extremum values of the undulator parameter,
$\pmin$ and $\pmax$, having been used in  (\ref{dE.4a}), 
produce the lower, $\om_j^{(\min)}$, and the upper, $\om_j^{(\max)}$, bounds 
on $\om_j$.
In the limit  $p_{\max}^2 \ll 1$  there is a weak variation of the harmonics 
frequencies,  $\om_j^{(\min)}\approx \om_j^{(\max)}$. 
However, in the case $\pmax > 1$ the change in the undulator parameter
leads to the emission within the band 
$\Delta\om = \om_j^{(\max)}-\om_j^{(\max)}$ which can greatly exceed not
only the peak width $\Delta\om_j$ but also the interval between the 
neighbouring harmonics.
In the latter case the monochromaticity of the radiation will be
smeared out if one carries out the averaging procedure 
(\ref{Averaging.1}).

However, the change in the undulator parameter and the harmonics
frequencies is not the only impact caused by the amplitude variation.
As mentioned above, in a crystalline undulator the peak intensity
is defined by the maximum value of the factor $\calD_N(\eta,\kpd,\kpa)$,
which depends on the ratios $\kpd=L/\Ld$ and $\kpa=L/\La$ (see (\ref{dE.5})).
Typically, in the crystalline undulators based on $\E = 0.6\dots10$ GeV
positrons channeling in crystals (so far Si monocrystals and 
Si-Ge mixtures were used 
\cite{Uggerhoj2006_Connell2006,BaranovEtAl_2006})
the energy of emitted photons lies within the range 
$10^2\dots 10^3$ keV. 
For these energies of positrons and photons the attenuation length 
$\La\equiv \La(\om)$ is on the level of several cm \cite{Hubbel} 
and by far exceeds the positron dechanneling lengths (lying within 
$\Ld \approx 0.03\dots 0.7$ cm in straight channels \cite{SPIE2007}).
Therefore, assuming $\kpa=0$ in (\ref{dE.5}), one finds that
the peak value of the factor $\calD_N(\eta,\kpd,\kpa)$ 
(i.e., at $\eta=j$) in an 
undulator with fixed number of periods, $N=L/\lambda$, is given by
\begin{eqnarray}
\lim_{\La\gg L,\Ld} \calD_{N}(j,\kpd,\kpa)
=
2N^2
\,
{1 - (1 + \kpd)\ee^{-\kpd} \over \kp_{\rm d}^2}
\label{Averaging.2}
\end{eqnarray}
The  dechanneling length in a channel, which is periodically bent 
with the amplitude $a$,
is defined by equations (\ref{dE.6}) and (\ref{dE.7}).
Hence, for $a=a(y)$ the right-hand side of (\ref{Averaging.2}) (and, generally, 
of (\ref{dE.5}) as well)
becomes dependent on
$y$ since $\kpd = L/\Ld(C) \propto (1-C)^{-2}$ with $C\propto a(y)$. 
In turn, the variation of the factor $\calD_N(\eta,\kpd,\kpa)$ can strongly 
influence the averaged spectral-angular distribution (\ref{Averaging.1}).

\section{Periodic deformations in bulk
\label{Formalism}}

\begin{figure}[ht]
\centering
\includegraphics[clip,width=13cm,height=5cm]{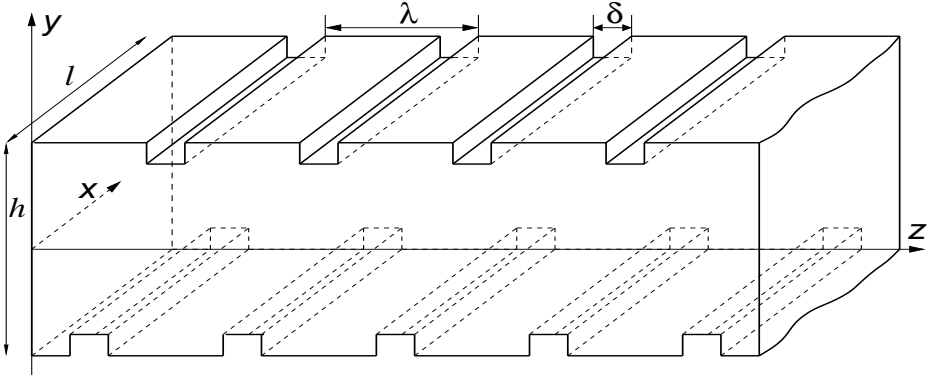}
\caption
{Schematic representation of a crystal with periodic surface 
deformations (the sets of regularly spaced grooves
parallel to the $x$ direction): 
$\lambda$ stands for the period of deformations and $\delta$ 
denote the width of a groove.
The set on the lower surface is shifted by $\lambda/2$ (along the 
$z$-axis) with respect to that on the upper surface.
The surface stress gives rise to the periodic bending of 
crystallographic planes in the bulk of crystal.
} 
\label{figure.2}
\end{figure}

In this section we present a formalism which allows one to 
carry out a quantitative analysis of the parameters of periodic
bending in the bulk of a crystal.
We consider the case when the bendings in the bulk are due to the 
periodic deformations on the crystal surfaces.
Such a situation is illustrated by Figure \ref{figure.2}:
two parallel opposite surfaces are deformed periodically  by means of
identical sets of parallel grooves applied to each of the surfaces
(the two sets are shifted by half-period, $\lambda/2$).

The deformation of this type can be achieved either by mechanical 
scratching of the crystal surface \cite{BellucciEtal2003,GuidiEtAl_2005} 
or by means of a more accurate laser-ablation method 
\cite{Uggerhoj2006_Connell2006}. 
Another possible method is in a deposition
of Si$_3$N$_4$ layers onto a Si wafer \cite{GuidiEtAl_2005}.

In either case the crystallographic planes in the bulk become
bent periodically although the shape $y=y(z)$ of bent planes 
does not follow  an ideal harmonic form $y=a\cos(2\pi z/\lambda)$.
The main deviations are: 
(a) the amplitude of bending depends on the distance from the surface,
and 
(b) in general case, higher subharmonics
(i.e. the Fourier components of $y(z)$ with smaller periods, 
$\lambda_n=\lambda/n$, where $n=2,3\dots$) contribute noticeably into 
the formation of the periodic shape.

However, as we demonstrate below, it is possible to establish the ranges of
parameters (these include $\lambda$, the thickness $h$ and elastic
constants of the crystal) within which the deviations of the resulting
periodic shape from the ideal form do not affect the spectral-angular
distribution of the undulator radiation.
  
\subsection{The equations of equilibrium with periodic boundary conditions
\label{Formalism1}}

To start with, let us formulate the approximations which will be 
used in our formalism.

Firstly, we assume that the width of a crystal $l$ (i.e. its size in the
$x$-direction, see figure \ref{figure.2}) is much larger than the 
thickness $h$: $l\gg h$.
As a result, one can disregard the deformations in the $x$-direction and
consider the deformations of a solid  body in the $(yz)$-plane only.  
Additionally, we assume that the period $\lambda$ is incomparably smaller than
the length $L$ of the crystal in the $z$ direction. 
Then, taking into account the periodicity of the surface deformations, 
one represents the displacement vector $\bfu(y,z)= (0,u_{y}(y,z),u_{z}(y,z))$ 
(i.e. the vector which characterizes the change in the position vector
of a point in the body due to the deformation) 
in the form of Fourier series:
\begin{equation}
\bfu(y,z)
= 
\sum_{n=-\infty}^{+\infty} \bfs_{n}(y)\, \ee^{\i n k z},
\label{Formalism1.1}
\end{equation}
where $k= {2 \pi/\lambda}$.
The vectors $\bfs_{n}(y)$ (with $n$ standing for an integer)
 are to be defined by solving the equation of equilibrium  
with proper boundary conditions.
In what follows we adopt that the $y$ coordinate is measured from the crystal
midline, and thus $y=-h/2$ corresponds to the lower surface, and
$y=h/2$ - to the upper one.

Secondly, we will consider the limit of small deformations only.
In this case the {\it strain tensor} $u_{ij}$ is defined as follows
(see, e.g., \cite{Landau7}):
\begin{equation}
u_{ij} 
= 
{1\over 2}
\left(
{\partial u_i \over \partial x_j} + {\partial u_j \over \partial x_i}
\right),
\label{Formalism1.2}
\end{equation}
with $i=x,y$ and $j=x,y$.

The third approximation concerns the {\it stress tensor}, $\sigma_{ij}$
(we remind, that physically $\sigma_{ij}$ stands for a stress on the $i$-th 
plane along the $j$-th direction).
In an isotropic media the components of the stress tensor can be related to 
$u_{ij}$ by means of the two elastic constants:  
the  Young's modulus $E$ and  the Poisson's ratio $\nu$.
In the case of a  planar deformation the relationship is as follows
(see, e.g., \S 5 in \cite{Landau7}):
\begin{eqnarray}
\cases{
\sigma_{yy} 
\displaystyle
= 
{E \over (1+\nu)(1-2 \nu)} 
\Bigl[(1-\nu) u_{yy} + \nu u_{zz}\Bigr], 
\\
\sigma_{zz}
=
{E \over (1+\nu)(1-2 \nu)} 
\Bigl[(1-\nu) u_{zz} + \nu u_{yy}\Bigr],  
\\
\sigma_{yz} 
= 
\sigma_{zy} 
= 
{E \over (1+\nu)} u_{yz}.
}
\label{Formalism1.3}
\end{eqnarray}
In anisotropic media (e.g., in a crystal) $E$ and $\nu$
depend on the directions of the applied stress and of the deformation.
For example, depending on a crystallographic direction 
the Poisson's ratio and the Young's modulus 
in a Si crystal varies within the intervals
$0.048<\nu<0.403$ and $130 < E < 170$ GPa \cite{WortmanEvans1965}. 
However, to simplify the analysis one can chose some average values.
In our numerical analysis we use $\nu=0.28$ and $E=150$ GPa which 
are close to the values used in 
modeling various deformation processes in silicon 
\cite{GuidiEtAl_2005,LuBogy_1995}.
To check the sensitivity of the results to the choice of the Poisson's ratio 
we also carried out the calculations using the extreme values of $\nu$ in
silicon.
It turned out that nearly an order of magnitude change 
in $\nu$ does not noticeably affect the results 
which are presented below in the paper.

The components of the stress tensor satisfy the following equations of 
equilibrium:
\begin{eqnarray}
{\partial \sigma_{yy} \over \partial y} 
+  
{\partial \sigma_{yz} \over \partial z} 
= 0, 
\qquad
{\partial \sigma_{zy} \over \partial y} 
+  
{\partial \sigma_{zz} \over \partial z} 
= 0 .
\label{Formalism1.4}
\end{eqnarray}
Using (\ref{Formalism1.1})--(\ref{Formalism1.3}) in (\ref{Formalism1.4}) 
one derives 
the system of coupled equations for the functions $s_{ny}(y)$ and $s_{nz}(y)$:
\begin{eqnarray}
\cases{
2(1-\nu) {\d^2 s_{ny}\over \d y^2}
+
\i n k \,{\d s_{nz}\over \d y}
- (nk)^2  
 \left(1-2\nu\right)
s_{ny}
= 0 
\\
\left(1-2\nu\right)
{\d^2 s_{nz}\over \d y^2}
+
\i nk\, {\d s_{ny}\over \d y}
- 2(1-\nu) (nk)^2 s_{nz}  
= 0.
}
\label{Formalism1.5}
\end{eqnarray}
To find the unique solution of this system  one has to impose the boundary
conditions. These can be formulated as follows.

Each trench acts as a source of a normal (`$\perp$') and a tangential, 
or a shear ($\|$) tension which are characterized by the average 
pressures $P^{\perp}$ and $P^{\|}$ applied to the crystal surface in 
vicinity of the trench.
In the case when the period $\lambda$ the width $\delta$
(see figure \ref{figure.2}), the trenches are equivalent to the 
sets of concentrated normal and shear forces applied along the equally 
spaced lines on the upper and lower surfaces
\footnote{The condition $\lambda\gg \delta$ is well fulfilled in the
 crystalline undulators which have been manufactured. 
Typical values are: $\lambda=100\dots500$ $\mu$m and $\delta\sim 10$ $\mu$m
\cite{BellucciEtal2003,GuidiEtAl_2005,Uggerhoj2006_Connell2006,BaranovEtAl_2006}.
}.
Within this model the pressures $P^{\perp}$ and $P^{\|}$ can be related to 
the components of the stress tensor calculated at the upper ($y=h/2$) and 
the lower ($y=-h/2$) surfaces.

Let us first formulate the boundary conditions due to the normal tension.
The pressure $P^{\perp}$ is applied inward the crystal along straight lines, 
parallel to the $x$ axis, passing through the equally-spaced points in the 
$z$ direction.
Therefore, recalling that $\sum_j \sigma_{ij}\mathfrak{n}_j$ represents 
the $i$-th 
component of the force per unit area (with $\mathfrak{n}_j$ standing 
for the $j$-th 
component of the outward-pointing normal) one derives the  boundary conditions:
\begin{eqnarray}
\cases{
\sigma_{yy}\Bigr|_{y=-h/2} 
= 
- \lambda P^{\perp} \sum_{n=-\infty}^{+\infty} \delta(z - \lambda n),
\\
\sigma_{yy}\Bigr|_{y= h/2} 
= 
- \lambda P^{\perp} 
\sum_{n=-\infty}^{+\infty} \delta\Bigl(z-\lambda (n+1/2)\Bigr), 
\\
\sigma_{yz}\Bigr|_{y=\pm h/2} = 0.
}
\label{Formalism1.6}
\end{eqnarray}
Here the coefficient $\lambda$ ensures that the period-averaged pressure, 
equals to $P^{\perp}$. 
The arguments of the delta functions fix the $z$ coordinates of 
the lines parallel to the $y$ axis.
As mentioned, the sets of trenches on the upper and lower surfaces are 
shifted by $\lambda/2$, and this explains the difference
of the arguments of the delta functions for  $y=h/2$ and $y=-h/2$.

In the case of shear stress, 
supposing that the trench profile is symmetric, one notices that the 
tangential forces created at the opposite edges of a trench are equal in 
magnitude but are antiparallel.
Therefore, the component $\sigma_{yz}$ as a function of $z$ must change the 
sign when crossing a trench.
It means that in the limit of infinitesimal width  $\delta\to 0$  this 
component becomes proportional not to the delta functions as in 
(\ref{Formalism1.6}) but to their derivatives.
As a result, one writes the boundary conditions in the form:
\begin{eqnarray}
\cases{
\sigma_{zy}\Bigl|_{y=-h/2} 
= 
F^{\|} \sum_{n=-\infty}^{\infty} \delta^{\prime}(z - \lambda n),
\\
\sigma_{zy}\Bigl|_{y= h/2} 
= 
- 
F^{\|}
\sum_{n=-\infty}^{\infty} \delta^{\prime}\Bigl(z - \lambda (n+1/2)\Bigr),
\\
\sigma_{yy}\Bigl|_{y=\pm h/2} = 0.
}
\label{Formalism1.7}
\end{eqnarray}
where $F^{\|}= \lambda^2 P^{\|}/2\pi$ stands for the tangential force 
associated with the period-averaged  pressure $P^{\|}$.

With the help of (\ref{Formalism1.1})--(\ref{Formalism1.3}) the equations
(\ref{Formalism1.6}) and (\ref{Formalism1.7}) can be re-written in terms of 
the functions $s_{ny}(y)$ and $s_{nz}(y)$ taken at $y=\pm h/2$.
The obtained formulae suffice to determine completely the solutions 
$s_{nj}^{\perp, \|}(y)$  ($j=y,z$) 
for  the normal or the shear stress. 
To obtain the solution in the case when both types of stress act simultaneously
one constructs the corresponding linear combinations of the functions
$s_{nj}^{\perp}(y)$ and $s_{nj}^{\|}(y)$.

Using these functions further in (\ref{Formalism1.1}) one determines the
displacement vector $\bfu(y,z)$.
The $y$ component of this vector is of a special interest in connection
of the crystalline undulator problem since it determines the profile of the
periodically bent channel in the bulk.

\subsection{Displacement $u_y(y,z)$ in the cases of normal and shear stresses
\label{Formalism2}}

Let us first analyze the periodic deformation in the bulk due to 
the normal and the shear stresses separately.

Resolving the system (\ref{Formalism1.5}) with the boundary conditions 
(\ref{Formalism1.6}) (or  (\ref{Formalism1.7})) one finds the functions 
$s_{ny}^{\perp}(y)$ (or $s_{ny}^{\|}(y)$) for all $n$.
Using these in (\ref{Formalism1.1}) one represents the 
$y$-component of the displacement vector in the following form: 
\begin{eqnarray}
u^{\perp,\|}_y(y,z)
=
-
\sum_{n=1}^{\infty} 
A_n^{\perp,\|}(y)\,\cos(n k z+\pi n).
\label{Formalism2.1}
\end{eqnarray}
Here $A_n^{\perp,\|}(y)$ can be associated with the amplitude of the 
$n$th harmonic of teh periodic bending,  i.e., the one with the period 
$\lambda_n=\lambda/n$.
A perfect crystalline undulator (see figure \ref{figure.1}, left panel) 
is characterized only by the term $n=1$ whose amplitude is independent on 
$y\in [-h/2, h/2]$. 
However, if a crystalline undulator is prepared by applying 
concentrated periodic normal or shear stress then:
(a) higher amplitude harmonics (with $n>1$) appear, 
and (b) the homogeneity of the bending amplitudes is lost since they 
become $y$-dependent. 
To stress these features of the amplitudes one can use the 
formulae which conveniently expresses $A_n^{\perp,\|}(y)$  via 
$A^{\perp,\|}_1(0)$ - 
the  first harmonic amplitudes is the midplane of the crystal 
(i.e. in the $(xz)$-plane with $y=0$).

In the case of normal stress alone the formula is as follows:
\begin{eqnarray}
\fl
A_n^{\perp}(y)
=
{A^{\perp}_1(0)\over n} 
{ \Delta_1^{-} \over \calI_1^{\perp} }
\times
\cases{
{\calI_n^{\perp} \cosh(nky) 
- 
2\calC_n\,n ky \sinh(nky)
\over \Delta_n^{-} },
& $n=1,3,5\dots$, 
\\
{
\calJ_n^{\perp} \sinh(nky)
-
2\calS_n\,nky \cosh(nky) 
\over \Delta_n^{+}
}
,
& $ n=2,4,6\dots$
}
\label{Formalism2.2}
\end{eqnarray}
For a shear stress the relationship reads:
\begin{eqnarray}
\fl
A_n^{\|}(y)
=
A_1^{\|}(0)\,
{\Delta_1^{-}\over \calI_1^{\|}}
\times
\cases{
{\calI_n^{\|}\cosh(nky) - 2\calS_n\,nky \sinh(nky)
\over  \Delta_n^{-}},
& $n=1,3,5\dots$, 
\\
{\calJ_n^{\|}\sinh(nky) - 2\calC_n\,nky \cosh(nky) 
\over  \Delta_n^{+}},
& $n=2,4,6\dots$
}
\label{Formalism2.3}
\end{eqnarray}

In (\ref{Formalism2.2}) and (\ref{Formalism2.3}) the following 
notations are used:
\begin{eqnarray}
\cases{
\Delta_n^{\pm} =  \sinh(2n\zeta) \pm 2n\zeta,
\\
\calC_n =\cosh(n\zeta),
\quad
\calS_n =\sinh(n\zeta),
}
\qquad
\zeta = {k h\over 2},
\qquad
k= {2\pi\over \lambda},
\label{Formalism2.5}
\end{eqnarray}
and
\begin{eqnarray}
\cases{
\calI_n^{\perp}
=
4(1-\nu)\calC_n + 2n \zeta \calS_n,
\\
\calJ_n^{\perp}
=
4(1-\nu)\calS_n + 2n \zeta \calC_n,
}
\qquad
\cases{
\calI_n^{\|}
=
2(1 - 2 \nu) \calS_n + 2n\zeta \calC_n,
\\
\calJ_n^{\|}
=
2(1 - 2 \nu) \calC_n + 2n\zeta \calS_n.
}
\label{Formalism2.7}
\end{eqnarray}
The amplitudes $A^{\perp,\|}_1(0)$ are proportional to 
the applied stresses and are given by:
\begin{eqnarray}
A_1^{\perp,\|}(0)
=
\lambda\,
{P^{\perp,\|}\over E} \,
\calF^{\perp,\|}(h/\lambda)
\label{Formalism2.8}
\end{eqnarray}
where
\begin{eqnarray}
\cases{
\calF^{\perp}(h/\lambda) 
=
{2(1+\nu)\over \pi}\,
{\zeta  \sinh(\zeta) + 2(1-\nu) \cosh(\zeta)  \over \sinh(2\zeta) - 2\zeta},
\\
\calF^{\parallel}(h/\lambda) 
=
{2(1+\nu)\over \pi}\,
{ (1-2\nu) \sinh(\zeta) + \zeta \cosh(\zeta) 
 \over \sinh(2\zeta) - 2\zeta}.
}
\label{Formalism2.8a}
\end{eqnarray}
Let us note that due to the half-period relative displacement of the
deformations on the upper and lower surfaces of the crystal
the amplitudes  $A_n^{\perp,\|}(y)$ with odd values of $n$ are even 
functions of $y$ and  vice versa
(see (\ref{Formalism2.2}) and (\ref{Formalism2.3})).
Therefore, it is sufficient to analyze the amplitudes in the upper half
of the crystal, i.e. for $y=[0, h/2]$.

\subsection{Numerical results for the bending amplitudes 
$A_n^{\perp}(y)$ and $A_n^{\|}(y)$ 
\label{Formalism3}}

\begin{figure}[t]
\vspace*{1cm}
\centering
\includegraphics[scale=0.64]{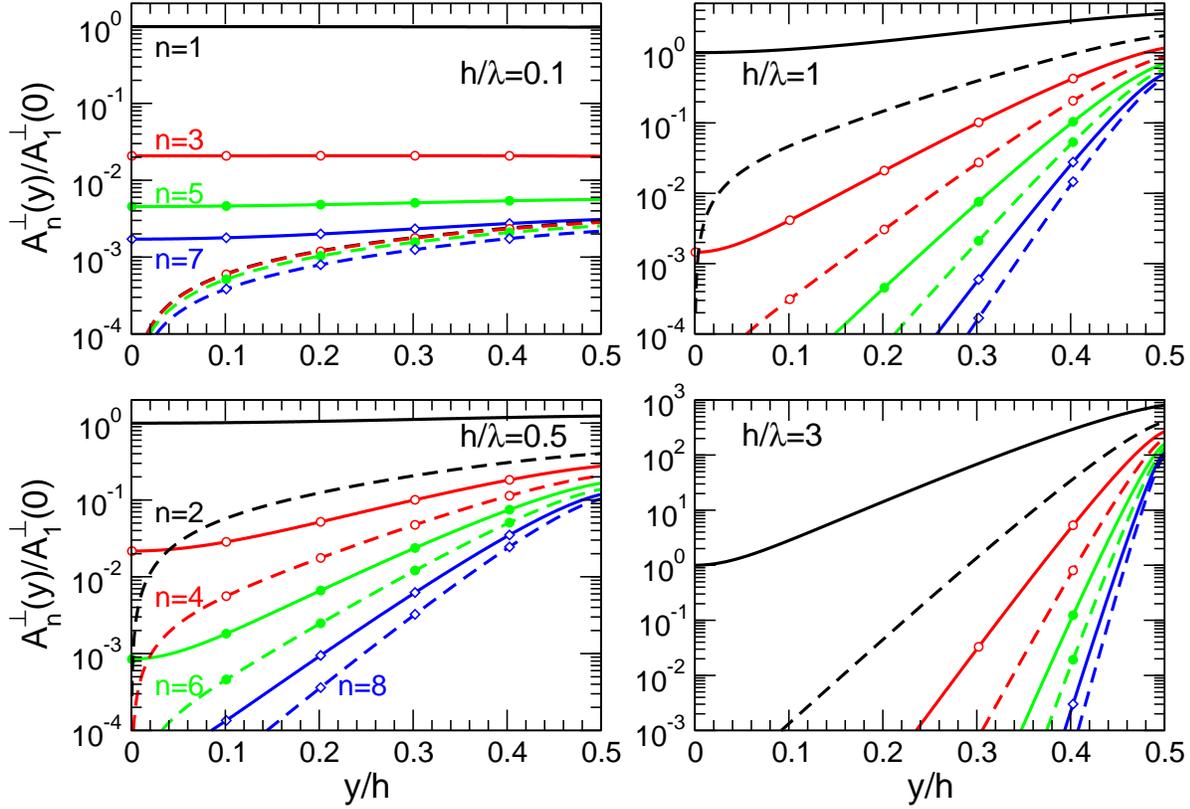}
\caption{ 
The ratio $A_n^{\perp}(y)/A_1^{\perp}(0)$
(see (\ref{Formalism2.2}) and (\ref{Formalism2.8})) 
versus $y/h$  (for $y\in[0,h/2]$) calculated 
for several values of the Si crystal thickness 
as indicated by the parameter $h/\lambda$.
The solid curves correspond to $n$ odd    
(which are indicated explicitly in the top left graph), 
the dashed curves - to $n$ even (see the bottom left graph).   
}
\label{figure.Any2A10_perp}
\end{figure}

\begin{figure}[ht]
\vspace*{1cm}
\centering
\includegraphics[scale=0.64]{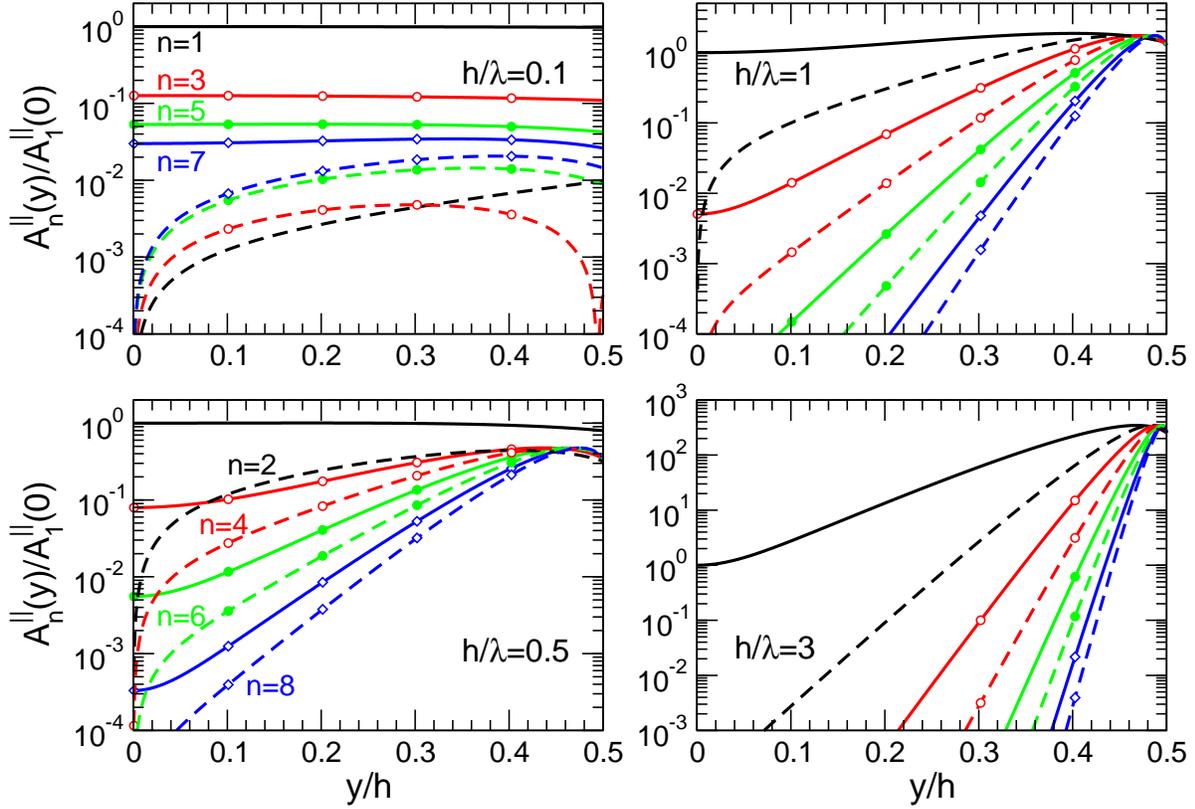}
\caption{ 
Same as in figure \ref{figure.Any2A10_perp} but for 
the ratio $A_n^{\|}(y)/A_1^{\|}(0)$
(see (\ref{Formalism2.3}) and (\ref{Formalism2.8})). 
}
\label{figure.Any2A10_par}
\end{figure}

The behaviour of the ratios
$A_n^{\perp}(y)/A_1^{\perp}(0)$ and $A_n^{\|}(y)/A_1^{\|}(0)$
as functions of the scaled distance $y/h$ from the midplane 
is illustrated by figures 
\ref{figure.Any2A10_perp} and \ref{figure.Any2A10_par}.
The dependences were calculated for several values of the crystal 
thickness $h$ (measured in the periods $\lambda$ as specified by 
the ratio $h/\lambda$) and several even and odd values of $n$.
The data refer to a Si crystal and the Poisson's ratio was chosen 
as $\nu=0.28$.


It is clearly seen from the figures that for either type of stress 
the inhomogeneity of the bending amplitudes along the
$y$ direction is much more pronounced for a thick crystal
(i.e., when $h > \lambda$) than for a thin one with $h < \lambda$. 

Indeed, in the limit $h \ll \lambda$ the amplitudes corresponding 
to odd values of $n\lesssim \lambda/h$ do not vary noticeably over 
the crystal thickness.   
Being even functions of $y$, the odd-$n$ amplitudes   $A_n^{\perp,\|}(y)$
depend quadratically on $y$ in the vicinity of the midplane,
$A_n^{\perp,\|}(y) \sim a + b\,n^2(y/\lambda)^2$, and have minimum 
at $y=0$.
In a thin crystal the quadratic term is small for all $y$ provided $n$
satisfies the condition written above.
As a result, $A_n^{\perp,\|}(y) \approx const$ for these $n$. 
The main reason for the even-$n$ amplitudes to vary even in the case of a thin
crystal is that they are odd functions of $y$.
Therefore, being non-zero at the surfaces $y=\pm h/2$ these amplitudes
attain zero in the midplane. 
(A non-monotonous behaviour of $A_n^{\|}(y)$ for particular $n$ and 
$h/\lambda$,
which is most pronounced for the  $n=4,6,8$ curves on the top left 
graph of figure \ref{figure.Any2A10_par}, 
just reflects the fact that the right-hand side of (\ref{Formalism2.3})
is not, generally, a monotonous function of $y$.)

Another important feature of a thin crystal is that over the whole
thickness the amplitude with $n=1$ greatly exceeds those with 
higher $n$.
This dominance is more pronounced for a normal stress
than for a shear one as reflected by additional factor $n^{-1}$ 
on the right-hand side of (\ref{Formalism2.2}).
As a result, the terms with $A_1^{\perp,\|}(y)\approx A_1^{\perp,\|}(0)$
prevail in the series from (\ref{Formalism1.1}), so that the periodic
bending in a thin crystal is of nearly harmonic shape:
$u^{\perp,\|}_y(y,z)\approx A_1^{\perp,\|}(0)\,\cos(2\pi z/\lambda)$.

As the crystal thickness increases the variation of the amplitudes
over half-width evolves dramatically. 
In the case of a thick crystal this variation reaches orders of magnitudes.
Such a behaviour one understands analyzing the right-hand sides 
of (\ref{Formalism2.2}) and (\ref{Formalism2.3}).
Assuming $h\gg \lambda$ 
one derives the following expressions for the amplitudes
on the surface and in the midplane:
\begin{eqnarray}
\fl
\begin{array}{lll}
\displaystyle{
A_n^{\perp}(h/2)
=
{b \over n}\, 
{\ee^{\zeta}\over \zeta}\,A^{\perp}_1(0),}
&
\quad
\displaystyle{
A_n^{\|}(h/2)
=
b\,{\ee^{\zeta}\over \zeta}\,A_1^{\|}(0),
}
&
\quad
n=1,2,3,4\dots
\\
A_n^{\perp}(0)
=
\ee^{(1-n)\zeta}\,A_1^{\|}(0),
&
\quad
A_n^{\|}(0)
=
n\,\ee^{(1-n)\zeta}\,A_1^{\|}(0),
&
\quad
n=1,3,5,\dots
\end{array}
\label{Formalism3.1}
\end{eqnarray}
where $b=(1-\nu)$ in the case of a normal stress and $b=(1-2\nu)/2$ for 
a shear stress.
For even $n$ the identity $A_n^{\perp,\|}(0)=0$ is valid.

Equations from (\ref{Formalism3.1}) demonstrate that for all $n$
the amplitudes decrease exponentially with the penetration distance
into the crystal.
For a fixed odd $n$ the decrease rate can be 
characterized by the ratio 
${A_n^{\perp,\|}(h/2)/A_n^{\perp,\|}(0)}= b\,\ee^{n\zeta}/n\zeta$, 
which is, basically,  independent on the type of applied stress 
(the difference manifests itself only in the pre-factor $b$).
On the other hand, the first line in (\ref{Formalism3.1}) indicates
that in the limit of a very thick crystal the amplitudes weakly 
depend on $n$ in a surface layer of the width $\sim \lambda$ 
(this dependence is more pronounced in the case of a normal stress due to 
the additional factor $n$).
These two features of the amplitude suggest 
that deviation of the periodic bending from the harmonic shape is
very strong in the outer layers of the crystal whereas in the central 
layer the terms with $n>1$ are negligibly small and the profile
of bending 
$u^{\perp,\|}_y(y,z)\approx A_1^{\perp,\|}(y)\,\cos(2\pi z/\lambda)$
is nearly perfect.
The deviation of the amplitudes $A_1^{\perp,\|}(y)$ from their values at
$y=0$ is described by the following formulae:
\begin{eqnarray}
\fl
A_1^{\perp,\|}(y)
\approx 
A^{\perp,\|}_1(0)
\left(
1 -
\varkappa^{\perp,\|}\,{(ky)^2 \over 2}
\right),
\quad
\mbox{where}
\quad
\cases{
\varkappa^{\perp}
=
{
2\nu  - \zeta \tanh(\zeta)
\over 
2(1-\nu) + \zeta \tanh(\zeta)
},
\\
\varkappa^{\|}
=
{
1 + 2 \nu  - \zeta \tanh^{-1}(\zeta)
\over  
1 - 2 \nu  + \zeta \tanh^{-1}(\zeta)
}.
}
\label{Formalism3.2}
\end{eqnarray}

To conclude this section we note that figures 
\ref{figure.Any2A10_perp} and \ref{figure.Any2A10_par}
suggest that the use of crystalline undulators manufactured by means of 
periodic surface deformation can be justified in two cases
 (irrelevantly to the type of the stress applied to crystal).

Firstly, it is the limit of a thin crystal $h < \lambda$ which ensures
(a) the (nearly) constant value of the amplitude $A_1(y) \approx A_1(0)$, 
and, 
(b) a small contribution of higher-$n$ terms to sum on 
the right-hand side of (\ref{Formalism2.1}).

Secondly, for $h>\lambda$ it is meaningful to use only the central
part of the crystal, i.e., where $|y| < \lambda/2\pi$, as an undulator.
In this case it is necessary to use a narrow beam (along the $y$-direction)
which is accurately aligned with the crystal midplane.

\subsection{Estimation of the variation of the undulator parameter $p$
and the bending parameter $C$
\label{Formalism4}}

Equation (\ref{Formalism2.1}) shows that in general case the 
profile of bending contains contributions of the terms with various 
$n$ thus deviating from a pure harmonic shape $\propto \cos(k z)$.
Hence, it is meaningful to analyze the influence of this deviation 
on the undulator parameter $p$ and the bending parameter $C$ which, as it 
was discussed at the end of section \ref{Averaging}, are of importance for 
the calculation of the averaged spectral-angular distribution  
(\ref{Averaging.1}) of radiation.

\subsubsection{Variation of the undulator parameter $p$.}
Generally, the parameter $p$ characterizes the mean-square velocity, 
$\overline{v_{\perp}^2}$, of the periodic 
transverse motion of a particle moving in an undulator (e.g., \cite{Baier}).
For an ultra-relativistic projectile the relationship is as follows: 
$p^2 = 2\gamma^2 v_{\perp}^2/c^2$.
In a perfect undulator this formula leads to equation (\ref{No.3}).
In the imperfect undulator discussed here, a particle moves along the
trajectory defined by the right-hand side of (\ref{Formalism2.1}) with
 $z\approx ct$. 
Calculating the transverse velocity $\d u_y(y,z)/ \d t$ and 
averaging over the period $T=\lambda/c$, one derives the 
expression for the undulator parameter as a function of $y$:
\begin{eqnarray}
p^2(y) = \sum_{n=1}^{\infty}  p_n^2(y),
\label{Formalism4.1}
\end{eqnarray}
where $p_n(y)= 2\pi\gamma {nA_n(y) / \lambda}$ stands for  the partial
undulator parameters corresponding to the $n$-th subharmonics of the 
trajectory. 

\begin{figure}[ht]
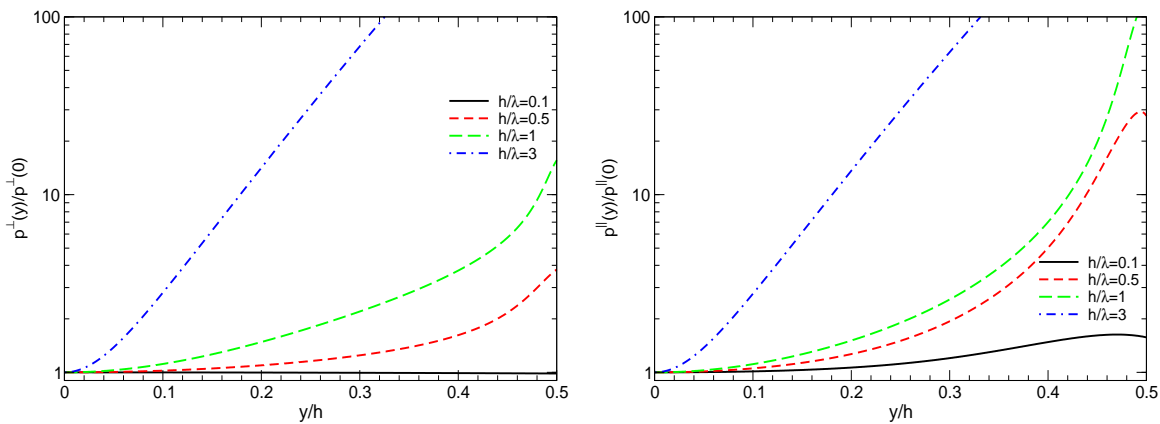

\vspace*{1cm}
\parbox{15.5cm}{
\includegraphics[width=7.5cm,height=5.5cm]{figure5a.eps}
\hspace*{0.2cm}
\includegraphics[width=7.5cm,height=5.5cm]{figure5b.eps}
}
\caption
{
Dependences  $p(y)/p(0)$ (see (\ref{Formalism4.1})) versus $y/h$
calculated for the normal (left) and shear (right) stress.
Different curves corresponds to several values of Si crystal thickness 
as indicated by the parameter $h/\lambda$.
} 
\label{figure.py}
\end{figure}

In figure \ref{figure.py} we present the dependences of the ratio 
$p(y)/p(0)$ on $y/h$ calculated for several values of the crystal thickness
and for the two types of stress, as indicated in the caption. 
As it was discussed in section \ref{Averaging} the dependence
of the undulator parameter on $y$ leads to a variation in the 
harmonic frequencies $\om_j$ (\ref{dE.4a}) over the crystal thickness.
If this variation  becomes large enough (e.g., when the shift in 
the harmonic position becomes comparable with the separation 
$\om_{j+1}-\om_j$ of the neighbouring harmonics) then the spectral-angular
distribution will loose the peak-like pattern typical for the 
undulator radiation.
From this viewpoint the graphs from figure \ref{figure.py} allow one to
estimate the  influence of the change in the undulator parameter 
across the crystal cross section on the shape of 
spectral-angular distribution.

For both types of stress the dependences $p(y)$ exhibit the common trend:
the undulator parameter slowly varies with $y$ in the limit of a 
thin crystal but becomes a rapidly (exponentially) 
increasing function for a thick crystal.
This is a direct consequence of the definition (\ref{Formalism4.1})
and the behaviour of the amplitudes $A_n^{\perp,\|}(y)$
(see  figures \ref{figure.Any2A10_perp} and \ref{figure.Any2A10_par}).
Another peculiarity, which one notices comparing the two panels in 
figure \ref{figure.py}, is that for the same value of $h/\lambda$ 
the ratio $p^{\|}(y)/p^{\|}(0)$ grows faster than $p^{\perp}(y)/p^{\perp}(0)$.
Such a feature, which is more pronounced for higher $h/\lambda$ values,
can be understood if one recalls the differences in the $n$-dependences
of the bending amplitudes in the case of normal and shear stresses,
and the behaviour of $A_n(y)$ as functions of $y$.
Indeed, 
for moderate and large $h/\lambda$ ratios and for 
both types of the stress, the amplitudes $A_n(y)$ with $n>1$, 
being rapidly increasing functions, satisfy the relations 
$A_n(0) \ll A_1(0)$ and  $A_n(h/2) \lesssim A_1(h/2)$ 
(see (\ref{Formalism3.1})).
Hence, for $y=0$ the sum  on the right-hand side of (\ref{Formalism4.1})
is defined, basically, by the term with $n=1$, whereas for 
larger $y$ also the terms with higher $n$  contribute noticeably to the sum.
On the other hand, 
it follows from (\ref{Formalism2.2}) and (\ref{Formalism2.3})  that 
$A_n^{\|}(y)/A_n^{\perp}(y)\propto n$, resulting in a similar estimate 
for the ratio of the partial undulator parameters:
$p_n^{\|}(y)/p_n^{\perp}(y)\propto n$.
Therefore, in the case of the shear stress the undulator parameter 
varies with $y$ more rapidly  than for the normal stress.

Typically, in crystalline undulators prepared by means of surface 
deformation and based on $\E = 0.6\dots10$ GeV positron channeling 
\cite{Uggerhoj2006_Connell2006,BaranovEtAl_2006}),
the values of parameter $p$ lie within the range $\sim 0.1\dots3$. 
The curves in figure \ref{figure.py} allow us to estimate the degree
of consistency of using the undulators with various $p\equiv p(0)$ 
and $h/\lambda$ values.

It was pointed out in section \ref{Averaging}, 
that there will be no dramatic change in the spectral-angular distribution
of the radiation due to the imperfectness of the undulator
provided the variation of the undulator parameter does not lead to
a noticeable change in the position of the peaks located at 
the frequencies $\om_j$ (see (\ref{dE.4a})).
To analyze this condition let us consider, for the sake of clarity, 
emission in the forward direction ($\theta=0$). 
In this case the frequency of the $j$-th harmonics, radiated from the 
the channel located at the distance $y$ from the midplane, is given by
 \begin{eqnarray}
\om_j(y) = {4 \gamma^2\omega_0\, j \over p^2(y) + 2}\,.
\label{dE.4b}
\end{eqnarray}
It is clear that in the limit $p(0)\ll 1$ the position of the peak will
be practically unchanged for those $y$ where $p^2(y)\ll 1$.
These two inequalities, depending on the absolute value of $p(0)$, 
allow for a wide-range variation of the ratio $p(y)/p(0)$, up to the 
order of magnitude.
It follows from figure \ref{figure.py}  that for a thin crystal 
such a situation can be realizes over the full thickness $h$.
In a thick crystal ($h>\lambda$) only the central part ensures the needed
variation of the undulator parameter.

In the opposite limit of large undulator parameters, when $p^2(0)\gg 1$,
the stability of the peak position can be achieved only in that part  
of the crystal where $p(0)\Bigl( p(y)-p(0)\Bigr) < 1$.
Taking into account rapid variation of the ratio $p(y)/p(0)$ in the 
case of a thick crystal, one deduces that this inequality can be ensured 
only by using very thin crystals with $h\ll \lambda$. 

\subsubsection{Variation of the bending parameter $C$.}
The channeling process in a bent crystal takes place  
if the centrifugal force, $\E/R$, due to the channel bending 
is less than the interplanar force $\dUmax$ \cite{Tsyganov}.
In a perfect crystalline undulator this condition,  applied
to the points of maximum curvature, results in the inequality 
(\ref{dE.6}) which relates $\dUmax$ with the period and the amplitude of 
periodic bending \cite{KSG1998}.
In the case when the profile of periodic bending contains a number of 
subharmonics, e.g., as in (\ref{Formalism2.1}), is it more constructive to 
relate the parameter $C$ to the mean-square curvature
$\overline{R^{-2}}$
where the averaging is carried out over the period $\lambda$.
Recalling that the curvature $R^{-1}$ is proportional to the modulus of 
the second derivative of the bending profile, $|\d^2 u_y(y,z)/ \d z^2|$,
and carrying out the averaging, one derives the following  
expression for average bending parameter as a function of $y$:
\begin{eqnarray}
\overline{C(y)} = \left(\sum_{n=1}^{\infty}  \overline{C_n^2(y)}\right)^{1/2},
\label{Formalism4.2}
\end{eqnarray}
where 
\begin{eqnarray}
\overline{C_n^2(y)}
= 
{1\over 2}\left({4\pi^2\E n^2 A_n(y)\over  \dUmax \lambda^2}\right)^2
\label{Formalism4.3}
\end{eqnarray}
is the mean-square partial bending parameter corresponding to the 
$n$-th subharmonics of the trajectory (\ref{Formalism2.1}). 

The parameter $\overline{C(y)}$, being  used in (\ref{dE.7}), defines
the dechanneling length $\Ld(C)$ as a function of $y$.
As mentioned in section \ref{Averaging},
the variation of $\Ld(C)$ over the crystal thickness
may influence destructively the averaged spectral-angular 
distribution (\ref{Averaging.1}).

\begin{figure}[ht]
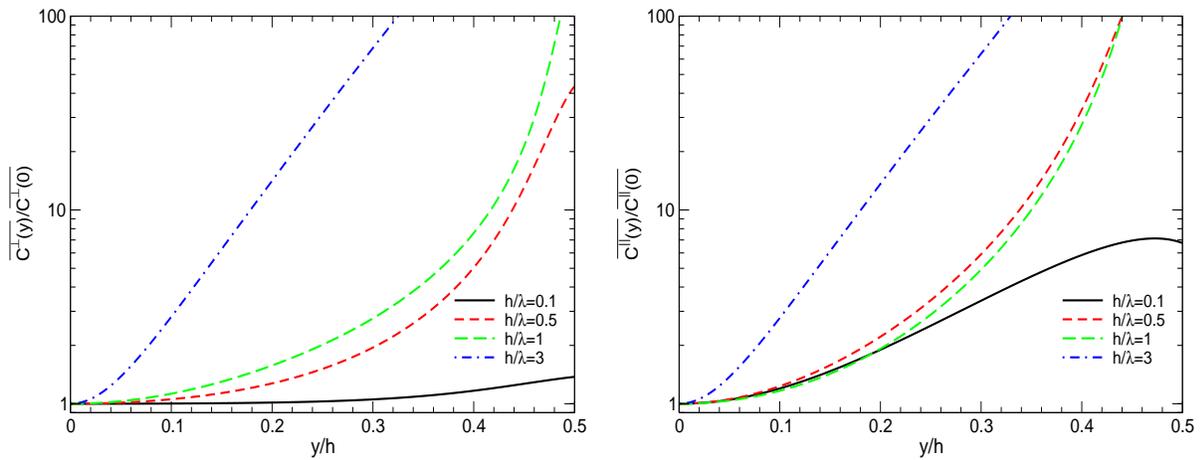

\vspace*{1cm}
\noindent
\parbox{15.7cm}{
\includegraphics[width=7.7cm,height=6cm]{figure6a.eps}
\hspace*{0.2cm}
\includegraphics[width=7.7cm,height=6cm]{figure6b.eps}
}
\caption
{
Ratios $\overline{C(y)}/\overline{C(0)}$ 
versus $y/h$
calculated for several values of the Si crystal thickness 
(as indicated by the parameter $h/\lambda$).
The left panel presents the dependences obtained for the normal stress.
The right panel -- for the shear stress.
} 
\label{figure.Cy}
\end{figure}

Figure \ref{figure.Cy} presents the dependences of $\overline{C(y)}$,
scaled by its value at the crystal center, versus $y/h$ 
calculated for several values of the crystal thickness
and for the two types of stress, as indicated in the caption. 
Qualitatively, the behaviour of the 
$\overline{C^{\perp,\|}(y)}/\overline{C^{\perp,\|}(0)}$ curves 
is similar to those of the ratio $p^{\perp,\|}(y)/p^{\perp,\|}(0)$ 
discussed above.
Nevertheless, there is a quantitative difference: for 
a fixed value of $h/\lambda$ the curves
$\overline{C^{\perp,\|}(y)}/\overline{C^{\perp,\|}(0)}$
increase faster with $y$ than the corresponding ratio of the undulator 
parameters.
This feature is due to the difference in the $n$-dependence of 
the partial terms in (\ref{Formalism2.1}) and in (\ref{Formalism2.2}).
Indeed, the partial bending parameters behave as
$\sqrt{\overline{C_n^2(y)}}\propto n^2 A_n(y)$
(see (\ref{Formalism2.3})) instead of the proportionality 
to $n A_n(y)$ of the partial undulator parameters.
It was discussed in connection with figure \ref{figure.py} that for 
moderate and large $h/\lambda$ values the terms with $n>1$ provide 
the increase of  $p(y)$ with $y$.
This increase is more pronounced for $\overline{C(y)}$ since its partial
terms contain an extra factor $n$. 
  
A crystalline undulator can operate only in the regime when $C<1$.
If otherwise, then the centrifugal force will drive the particles out of 
the channel.
More detailed analysis \cite{Dechan01} indicated that the reasonable range
for the bending parameter is $0.01\dots0.3$.
It means that if for $y=0$ the bending parameter is of the order of 
$10^{-2}$, then the variation of  $\overline{C(y)}$ within the order of 
magnitude are acceptable.
From figure \ref{figure.Cy} it follows that for a thin crystal 
such a situation can be realizes over the full thickness.
In a thick crystal ($h>\lambda$) only the central part ensures the needed
variation of the bending parameter.

\subsection{Calculation of $\lambda$ corresponding to given amplitudes at the 
crystal center \label{Formalism5}}

The numerical data, discussed above in sections \ref{Formalism3} and
\ref{Formalism4}, represent the $y$-dependence of the amplitudes, 
undulator parameters and bending parameters {\it scaled} by their
values at $y=0$.
The latter, in turn, can be expressed in terms of the amplitudes 
 $A_1^{\perp,\|}(0)$ which can be calculated from 
(\ref{Formalism2.8}) and (\ref{Formalism2.8a}). 
These equations relate the amplitudes to the bending period $\lambda$.

In connection with a perfect crystalline undulator it was established
(for a detailed discussion see \cite{KSG2004_review})
that the operation of the undulator should be considered in 
the large-amplitude regime, i.e. when the bending amplitude is much larger
than the interplanar distance $d$. 
In this limit the characteristic frequencies of undulator and 
channeling radiation (see, e.g., Ref.~\cite{Kumakhov2}) 
are well separated.
As a result, the channeling radiation does not affect the parameters of the 
undulator radiation, whereas the intensity of undulator radiation becomes
comparable or higher than that of the channeling one \cite{KSG1999}.

To apply this approach to a crystalline undulator with varied amplitude
one assumes that the large-amplitude regime is applicable to the 
amplitudes at the crystal midplane: $A_1^{\perp,\|}(0) > d$.
For the convenience of further consideration let us introduce 
the quantity
\begin{eqnarray}
\alpha^{\perp,\|} = {A_1^{\perp,\|}(0)\over d} > 1,
\label{Formalism5.1}
\end{eqnarray}
which explicitly measures the amplitude in the units of interplanar 
separation.

Other quantities, which enter equations 
(\ref{Formalism2.8}) and (\ref{Formalism2.8a}) include:
\\
(a) The crystalline medium dependent parameters,
the  Poisson's ratio $\nu$ and the Young's modulus $E$.  
(As already mentioned, for a Si crystal we use the average values 
$\nu=0.28$ and $E=150$ GPa.)
\\
(b) The crystal thickness $h$, which enters via the ratio 
$h/\lambda$.
\\
(c) The applied stress, $P^{\perp}$ or $P^{\|}$.

To estimate the stress one takes into account that we
are interested in elastic deformations of the crystalline structure.
Therefore, $P^{\perp,\|}$ must not exceed the plastic yield strength, $Y$,
which stands for the stress at which material strain changes 
from elastic deformation to the plastic one.
For a silicon crystal one can adopt $Y=7$ GPa \cite{YounKang2004}.
For further use let us introduce the quantity, which stands 
for the stress measured in the units of $Y$:
\begin{eqnarray}
\kappa^{\perp,\|} = {P^{\perp,\|}\over Y} \leq 1.
\label{Formalism5.2}
\end{eqnarray}

Using (\ref{Formalism5.1}) and (\ref{Formalism5.2}) 
one re-writes equation (\ref{Formalism2.8}) as follows
(to simplify the notations the superscripts `$\perp$' 
and/or `$\|$' are omitted):
\begin{eqnarray}
\lambda
=
{\alpha \over \kappa}\,
{E\over Y} \,
{d\over \calF(h/\lambda)}. 
\label{Formalism5.3}
\end{eqnarray}
Explicit forms of the functions $\calF(h/\lambda)$ for the two types of 
stress are given in (\ref{Formalism2.8a}).

Relationship (\ref{Formalism5.3}) allows one to find the values of 
$\lambda$ and $h$ which ensure, for a given crystal (the parameters $E$, 
$\nu$ and $d$) and for a relative stress (the parameter $\kappa$),
a desired value of the relative amplitude $\alpha$ in the center of crystal.

\begin{figure}[ht]
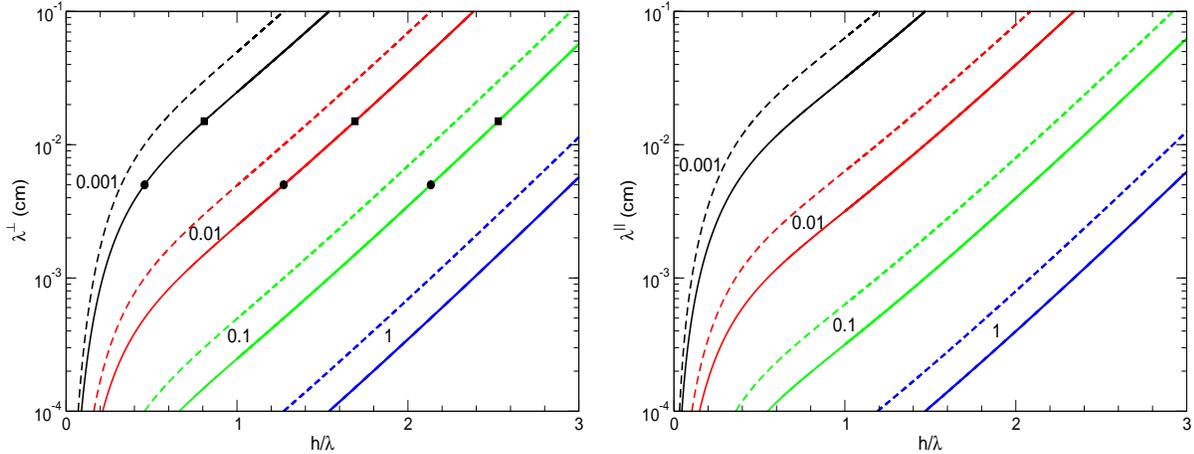

\vspace*{1cm}
\noindent
\parbox{15.7cm}{
\includegraphics[width=7.7cm,height=6cm]{figure7a.eps}
\hspace*{0.2cm}
\includegraphics[width=7.7cm,height=6cm]{figure7b.eps}
}
\caption
{
Dependence of $\lambda$ on the relative thickness $h/\lambda$  
(see (\ref{Formalism5.3}))
for several values of $\kappa = {P^{\perp,\|}/ Y}$ (as marked).
For each $\kappa$ the solid curve corresponds to $A_1^{\perp,\|}(0)/d=10$, 
the dashed ones - to $A_1^{\perp,\|}(0)/d=20$ 
(with $d=1.92$ \AA\, being the distance
between the (110) planes in Si).
Left panel presents the dependences obtained for the normal stress,
right panel -- for the shear stress.
The filled circles and squares on the left panel mark the values of 
$\lambda$ and $h$ for which the averaged spectra (\ref{Averaging.1})
were calculated (see section \ref{AveragedSpectra} for the details).
} 
\label{figure.lambda}
\end{figure}

To illustrate the relationship we present figure \ref{figure.lambda}, where
the dependence of $\lambda$ on $h/\lambda$ is plotted
for several values of $\kappa$ (as indicated) and for two values of 
the relative amplitude, $\alpha=10$ and $\alpha=20$. 
The two panels in the figure correspond to different types of the applied
stress.
It is seen that the $\lambda^{\perp}$ and $\lambda^{\|}$ 
curves obtained for the
same values of $\kappa$ and $\alpha$ look much alike although there is a 
distinguishable quantitative difference 
(note the log scale of the vertical axis).

\section{Averaged spectra \label{AveragedSpectra}}

In this section we present the results of numerical analysis of the influence
of the periodic bending imperfectness on the spectral-angular distribution.
The calculations were performed for two energies,
$\E=0.6$ GeV and $\E=5$ eV, of a positron channeling along periodically
bent (110) crystallographic planes in Si (the interplanar distance 
$d=1.92$ \AA,
the maximal interplanar force $\dUmax=6.35$ GeV/cm).
The data presented below refer to the emission in the forward direction.
(i.e., $\theta=0^{\circ}$ with respect to the $z$ axis, see figure 
\ref{figure.2}).
In this case the integrals (\ref{dE.3}) can be evaluated analytically and the 
spectral distribution (\ref{dE.1}) from an undulator with fixed parameters
is expressed in terms of the Appel function \cite{SPIE2007,Gradshteyn}.
These features, being not too important physically, simplify the numerical 
analysis.

The parameters of the crystalline undulators used in the calculations were 
chosen as follows.
Firstly, for both mentioned positron energies the amplitude $A_1(0)$ of the 
$n=1$ subharmonic in the crystal center was fixed as $\alpha = A_1(0)/d=10$, 
thus satisfying the large-amplitude condition (\ref{Formalism5.1}).
Secondly, for each $\E$ the length $L$ of the crystal  was chosen to be equal 
to the dechanneling length in the straight channel (see (\ref{dE.7})):  
$L = 0.041$ cm  and  $L = 0.31$ cm for $\E=0.6$ and $\E=5$ GeV, 
correspondingly.
The bending period is $\lambda=50\, \mu$m for  $\E=0.6$ GeV 
(resulting in $N=L/\lambda=8$ undulator periods), 
and $\lambda = 150\, \mu$m for  $\E=5$ GeV (with $N=20$).  
Using (\ref{dE.4b}) with $j=1$ one calculates the energies of the first 
harmonics emitted in the perfect undulators (i.e., with the cited $\lambda$ 
and $\alpha$ values):
$\hbar\om_1 = 65.5$ keV and $\hbar\om_1 = 1.2$ MeV for 
 $\E=0.6$ and $\E=5$ GeV, correspondingly.
Let us note that the mentioned values of 
positron energies, the crystal lengths and the parameters of periodically bent
channels are close to those discussed recently in connection with the 
experiments on crystalline undulators 
\cite{Uggerhoj2006_Connell2006,SPIE2007}.  

\begin{figure}[ht]
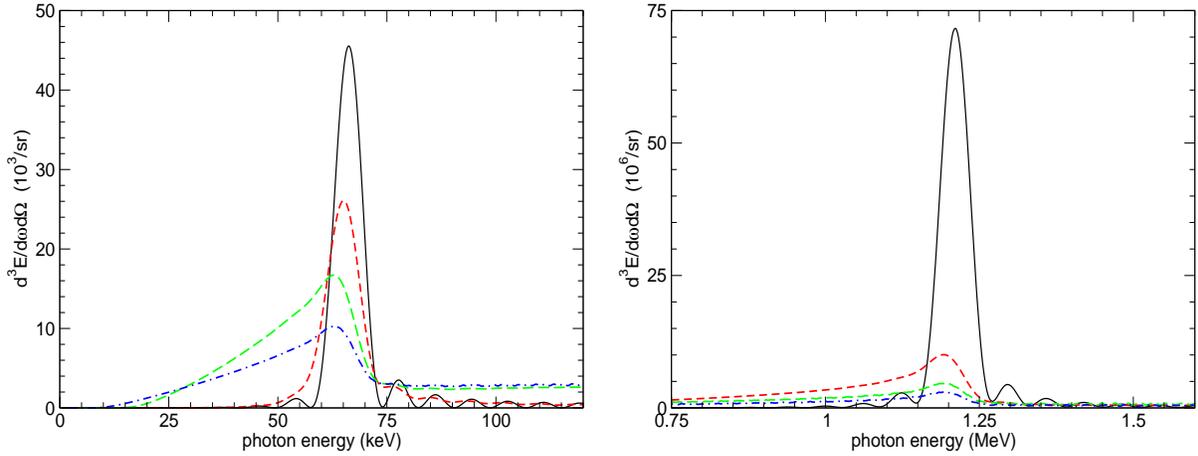

\vspace*{1cm}
\noindent
\parbox{15.7cm}{
\includegraphics[width=7.7cm,height=6cm]{figure8a.eps}
\hspace*{0.25cm}
\includegraphics[width=7.7cm,height=6cm]{figure8b.eps}
}
\caption
{
Spectral intensity of the undulator radiation emitted in the 
forward direction by a $0.6$ GeV (left panel) and 
a $5$ GeV  (right panel) positron channeling along
periodically bent  (110) planes in Si.
In each graph the solid curve stands for the intensity from a perfect
undulator with the fixed amplitudes $A_1(0)=10d$.
The dashed, long-dashed and chained curves present the averaged spectra
obtained for the crystals of different thickness $h$ and exposed to different
values of the normal stress.
Further explanations see in the text. 
}
\label{figure.dE}
\end{figure}

\begin{figure}[ht]
\vspace*{1cm}
\centering
\includegraphics[scale=0.45]{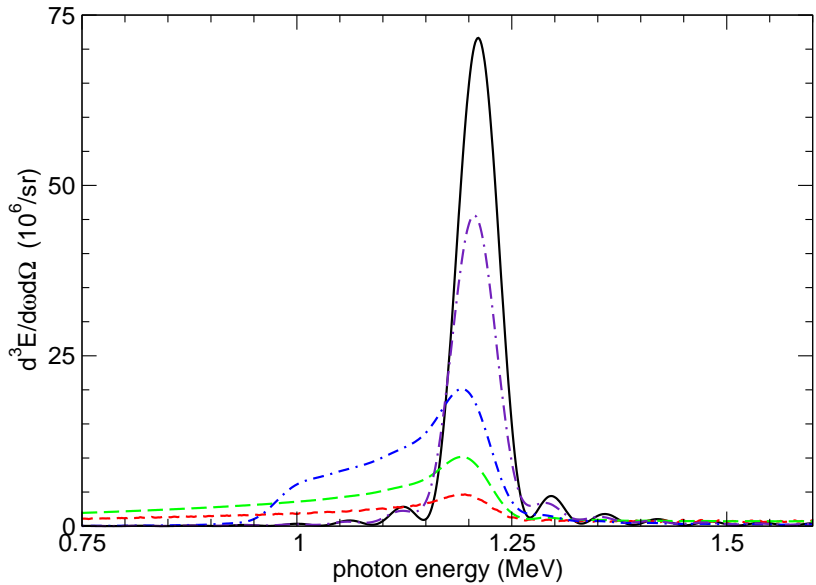}
\caption
{
Spectral intensity of the undulator radiation in the 
forward direction for a $5$ GeV  positron channeling along
periodically bent Si(110).
The undulator period, $\lambda = 150\, \mu$m, 
and the crystal thickness, $h=1.7\lambda$, 
correspond to the relative normal stress $\kappa^{\perp}=0.01$ 
(these parameters
are indicated in figure \ref{figure.lambda} by a central filled square).
The the solid curve stands for the intensity from a perfect
undulator with the fixed amplitude $A_1(0)=10d$.
Other curves represent the intensities (\ref{Averaging.1}) averaged
over different intervals $y=[0,y_{\max}]$:
the short-dashed curve corresponds to $y_{\max}=h/2$,
the long-dashed curve -- to $y_{\max}=0.2 h/2$,
the short-dashed chained curve -- to $y_{\max}=0.1 h/2 $,
the long-dashed chained curve -- to $y_{\max}=0.05 h/2$.
Further explanations see in the text. 
} 
\label{figure.Part_of_h}
\end{figure}

The results of calculations of the spectral distributions are presented in 
figures \ref{figure.dE} and \ref{figure.Part_of_h}.

Two graphs in figure \ref{figure.dE} correspond to two 
energies $\E$, as indicated in the caption.
In each graph, the solid curve represents the profile of the first harmonic 
peak (in the forward direction) calculated using (\ref{dE.1}) 
for the perfect undulators with the parameters $\alpha$ and $\lambda$ given
above. 
Other three curves in each graph correspond to the averaged spectra 
calculated for the same values of $\alpha$ and $\lambda$
but for different crystal thicknesses $h$. 
These spectra were obtained from (\ref{Averaging.1}) by setting 
$y_{\max}=h/2$ (the effective range of integration was restricted by the 
condition $\overline{C(y)}<1$, see the footnote comment below eq.
(\ref{Averaging.1})).
As it was mentioned in section \ref{Formalism5},  
  one can vary the crystal thickness together with
the relative stress  $\kappa$ to achieve fixed values of $\alpha$
(see (\ref{Formalism5.3})). 
The short-dashed, long-dashed and chained curves in the figure
were obtained for the relative stress 
$\kappa^{\perp}=0.001$, $\kappa^{\perp}=0.01$ and $\kappa^{\perp}=0.1$, 
respectively.
The corresponding values of $h$ one finds from figure \ref{figure.lambda} 
(left), where the filled circles mark the ratios $h/\lambda$ for $\E=0.6$ GeV
and the filled squares -- for $\E=5$ GeV. 

Comparing different curves in the figure and recalling the discussion
presented in section \ref{Averaging}, one sees the extent to which  
the imperfectness of the undulator structure over the crystal
thickness can influence the emission spectrum.
The increase in $h$ (or, more generally, in the interval over which the 
averaging in (\ref{Averaging.1}) is carried out) leads to a more
pronounced variation of the amplitudes $A_n(y)$ (see figures
\ref{figure.Any2A10_perp} and \ref{figure.Any2A10_par}).
This, in turn, trigger strong variations of the effective 
undulator parameter $p$ and bending parameter $C$, --
figures \ref{figure.py} and \ref{figure.Cy}.
The increase in $p$ with $y$ results in the decrease of the 
first harmonic energy (see (\ref{dE.4b})).
In the figure this feature is reflected by the (relative) enhancement
of the photons with the energy smaller than
the values of $\hbar\om_1$ cited above.
Thus, in the peaks in the averaged spectrum become wider and the width 
increases with $h$.
Another tendency,  clearly seen in the figure, is the decrease in the
peak intensity with $h$. 
This is mainly due to the variation of the bending parameter $\overline{C(y)}$.
Indeed, $C$ varies  from its minimum value $\overline{C(0)}$ 
at the center up to the
$\overline{C(h/2)}$ in the surface layer, figure \ref{figure.Cy}.
For a given $y$ the value $\overline{C(y)}$ defines the dechanneling
length $\Ld(\overline{C(y)})\propto (1-\overline{C(y)})^2$, 
which, in turn, influences the peak value of 
the spectrum via the factor $\calD_N(\eta,\kpd,\kpa)$: 
smaller $\Ld$ result in smaller peak intensities
(see (\ref{dE.5}) and section \ref{Averaging}).
Additionally, the channel acceptance (\ref{dE.8}) decreases with 
$\overline{C(y)}$ increasing.
Therefore, the relative contribution of the
trajectories with larger $\overline{C(y)}$-values to the integral from 
(\ref{Averaging.1}) increases with $h$ leading to the decrease 
in the peak intensity of the averaged spectrum.
 
The two features mentioned above are more pronounced for a positron energy
$\E=5$ GeV than for $\E=0.6$ GeV (compare right and left graphs in 
figure \ref{figure.dE}).
The reason is a s follows.
For a 5 GeV positron,  the undulator parameter at the center 
is $p(0) = 0.80$ which is noticeably larger than the value $p(0)=0.28$ for 
a $\E=0.6$ GeV positron.
Hence, the influence on the position of the first harmonic peak
due to the variation of the undulator parameter with $y$ is smaller for 
lower $\E$. 
As a result, the averaged peaks for $\E=5$ GeV are (relatively) wider 
than those for  $\E=0.6$ GeV.
The more pronounced decrease in the intensities for a $5$ GeV positron
is due to the larger contribution of the trajectories with higher 
$\overline{C(y)}$-values.
This, in turns, happens because 
for the same value of the relative stress $\kappa$
the relative thickness $h/\lambda$ is higher for 
a  $\E=5$ GeV  positron 
(compare the positions of filled squares 
and filled circles which correspond to the same $\kappa$ values
in figure (\ref{figure.lambda})).

Figure  \ref{figure.dE} demonstrates that the pattern of  
spectral-angular distribution can change dramatically. 
The well-separated peak-like structure, typical for the emission
spectrum from a perfect undulator, can be completely smeared out 
in the imperfect undulator due to the variation of the parameters of 
periodic bending.
The degree to which the peaks are destroyed depends on the crystal
thickness and on the values of the undulator parameter $p(0)$ and the bending
parameter $C(0)$ at the center of the crystal.
In particular, for $p(0)>1$ and in the case of a thick crystal 
($h\gg \lambda$) the averaged spectrum has a nearly uniform distribution 
which is typical for the noise rather than for the undulator-type radiation.

However, even in the case of a comparatively thick crystals one can
restore the coherence of radiation.
To achieve this, it is necessary to avoid using the layers of the
crystal located far off the centerline, i.e. to use 
not the whole thickness of the crystal but its central part of the 
width lower than
$\lambda$.
In the central layer, the amplitudes $A_n(y)$ and the related quantities 
do not deviate noticeably from their values at $y=0$ 
(see (\ref{Formalism3.2})), and, therefore, the averaging procedure 
will not radically influence the peak profile.

To illustrate this statement, we present figure \ref{figure.Part_of_h} which
contains the spectral intensity in the vicinity of the first harmonic  
for the perfect undulator (the solid line) along with the averaged spectra
calculated for different $y_{\max}\leq h/2$.
The data refer to $\E=5$ GeV positron channeling in a Si crystal with
$h/\lambda = 1.7$, the relative normal stress $\kappa^{\perp}=0.01$, 
the parameters $\alpha$ and $\lambda$ as in figure \ref{figure.dE} 
right.
The solid curves in figures \ref{figure.Part_of_h} and  \ref{figure.dE} (right)
are identical.
The short-dashed curve (the lowest one) in figure  \ref{figure.Part_of_h}
stands for the spectrum averaged over the whole thickness, i.e.
$y_{\max}= h/2$.
Other curves corresponds to smaller $y_{\max}$ values, as indicated.
It is clearly seen that by narrowing the averaging interval the profile
of the line can be made closer to that for the perfect undulator.

\section{Conclusions \label{Conclusion}}

In our paper we discussed the influence of imperfect structure of a 
crystalline undulator on the spectral distribution of the radiation.
We mainly analyzed the undulators in which the periodic bending in the
bulk appear as a result of regular surface deformations.
We demonstrated that this method inevitably leads
to two main deviations from the perfect harmonic shape $a\cos(2\pi z/\lambda)$.
Thsese are: (a) the dependence of the bending amplitude on the  
distance $y$ from crystal midplane, and 
(b) the presence of subharmonics with smaller bending periods. 
As a result, the quantities which characterize the crystalline undulator, 
-- 
the undulator parameter $p$ and the bending parameter $C$,
vary over the crystal thickness $h$.
In turn, this leads to the loss of the monochromaticity of the radiation 
formed in the undulator. 

Typical scale, within which the parameters vary noticeably, is equal
to the period $\lambda$ of the surface deformations.
As we demonstrated in the paper,
one can choose the following two strategies
to partly restore the monochromaticity of radiation.
Firstly, one can use thin crystals, $h<\lambda$. 
In this case, the variation of the amplitude over the 
width as well as the contribution of higher subharmonics do not lead to 
dramatic changes in the spectrum. 
However, this limit corresponds to very thin crystals, if one takes 
into account that the period of surface deformations lies within the 
range  $50\dots 200$ $\mu$m 
\cite{BellucciEtal2003,GuidiEtAl_2005,Uggerhoj2006_Connell2006}.

The second approach prescribes the use of a thick crystal but in combination
with a narrow ($\sigma_y\ll \lambda$) positron beam along the midplane.
This limit seems to be achievable by using existing positron beams in the GeV
range and with the size (along one direction) of several microns
(see \cite{ParticleDataGroup2006}). 

To minimize the destructive role of the imperfect structure,
one can also consider alternative schemes of the surface deformations.
Namely, if instead of just a periodic surface deformation one 
applies a {\em harmonic} surface deformation with a period $\lambda$, 
then the only imperfectness of the periodic bending in the bulk 
will be associated with the variation of the amplitude $A_1(y)$ since
for the amplitudes of subharmonics with $n\geq 2$ will be identically 
equal to zero.
This, in turn, will result in a much smaller variation of the
undulator parameter and the bending parameter over the crystal 
width (see eqs. (\ref{Formalism4.1}) and (\ref{Formalism4.2}), 
and figures \ref{figure.py} and \ref{figure.Cy}) since this variation
is to a great extent due to the contribution of higher subharmonics. 
To achieve the harmonic shape of the surface deformation one can
either place a crystal between two press molds of the harmonic profile
(shifted by $\lambda/2$ with respect to each other) or 
apply a modulated pressure by means of two piezoelectric layers.

\ack

The authors are grateful to Kim Kirsebom and Ulrik Uggerh{\o}j 
for constructive discussions.
This work has been supported by the European Commission 
(the PECU project, contract No. 4916).


\end{document}